# Additive interfacial chiral interaction in multilayers for stabilization of small individual skyrmion at room temperature

C. Moreau-Luchaire[1], C. Moutafis[2,3], N. Reyren[1], J. Sampaio[1,*], C. A. F. Vaz[2], N. Van Horne[1], K. Bouzehouane[1], K. Garcia[1], C. Deranlot[1], P. Warnicke[2], P. Wohlhüter[2,4], J.-M. George[1], M. Weigand[5], J. Raabe[2], V. Cros[1], A. Fert[1]

[1] Unité Mixte de Physique, CNRS, Thales, Univ. Paris-Sud, Université Paris-Saclay, 91767, Palaiseau, France.
[2] Swiss Light Source, Paul Scherrer Institute, Villigen, Switzerland.
[3] School of Computer Science, University of Manchester, Manchester, UK.
[4] Laboratory for Mesoscopic Systems, Department of Materials, ETH Zurich, Zurich, Switzerland.
[5] Max Planck Institute for Intelligent Systems, Stuttgart, Germany.
*present address : Laboratoire de Physique des Solides, CNRS, Univ. Paris-Sud, Université Paris-Saclay, 91405, Orsay, France.

Facing the ever-growing demand for data storage will most probably require a new paradigm. Nanoscale magnetic skyrmions are anticipated to solve this issue as they are arguably the smallest spin textures in magnetic thin films in nature. We designed cobalt-based multilayered thin films where the cobalt layer is sandwiched between two heavy metals providing additive interfacial Dzyaloshinskii-Moriya interactions, which reach a value close to 2 mJ m$^{-2}$ in the case of the Ir|Co|Pt asymmetric multilayers. Using a magnetization-sensitive scanning x-ray transmission microscopy technique, we imaged small magnetic domains at very low field in these multilayers. The study of their behavior in perpendicular magnetic field allows us to conclude that they are actually magnetic skyrmions stabilized by the large Dzyaloshinskii-Moriya interaction. This discovery of stable sub-100 nm individual skyrmions at room temperature in a technologically relevant material opens the way for device applications in a near future.

To process and store the continually increasing quantity of information constitutes a major challenge that is the target of many research programmes. The hard disk drives, in which information is encoded magnetically, allow the storage of zettabytes ($10^{21}$) of information, but this technology will soon reach its limits. An up-and-coming avenue is opened by the discovery of magnetic skyrmions[1,2], *i.e.* spin windings that can be localized within a diameter of a few nanometers and can move like particles[3]. These magnetic solitons, remarkably robust against defects due to their topology[4], are promising as being the ultimate magnetic bits to carry and store information in magnetic media. The topology of the skyrmions[5] also appears to further underline other important features such as their current-induced motion induced by small dc currents that is crucial for applications but also the existence of a specific component in Hall effect[6-8] that can be used for an electrical read-out of the information carried by nano-scale skyrmions[9]. We proposed recently that these skyrmions could be used in future information storage and processing devices[3].

The existence of the skyrmion spin configuration has been predicted theoretically about thirty years ago[1] but it was only recently that skyrmion lattices have been observed in crystals with non-centrosymmetric lattices, in particular in B20 structures such as MnSi[8,10-12], FeCoSi[13] or FeGe[7] crystals. Although skyrmions have been observed only below room temperature (r.t.) in such compounds, a very recent publication by Tokunaga *et al*[14] has reported the observation of skyrmion lattices above r.t. in *β*-type CuZnMn alloys. In 2011, skyrmions have also been identified in monolayers (Fe) or bilayers (Fe|Pd) of ferromagnetic metals with out-of-plane magnetization deposited on heavy metal substrates, Ir(1 1 1)[15-16]. Thin magnetic films appear to be more compatible with technological developments, although the observation of nanoscale skyrmions in thin films has been limited, until recently, to low temperatures[16]. The study of these new magnetic phases associated with chiral interactions generated a very strong interest in the community of solid-state physics.

The magnetic skyrmions, are in most cases, induced by chiral interactions of the Dzyaloshinskii-Moriya (DM) type, which result from spin-orbit effects in the absence of inversion symmetry. The hamiltonian of the DM Interaction (DMI) between two neighboring atomic spins $S_1$ and $S_2$ can be expressed as: $H_{DMI} = \mathbf{D}_{12}\cdot(\mathbf{S}_1\times\mathbf{S}_2)$ where $\mathbf{D}_{12}$ is the DM vector (top of Fig. 1a)[17]. In this work, we will focus on ultrathin magnetic systems in which the DMI results from the breaking of the inversion symmetry at the interface magnetic layers and heavy metals which provide a strong spin-orbit coupling, leading to DMI magnitudes nearly as large as the Heisenberg exchange[16]. In such a case, the DM vector $\mathbf{D}_{12}$ is perpendicular to the $\mathbf{r}(S_1)-\mathbf{r}(S_2)$ vector ($\mathbf{r}(\cdot)$ being the position vector) and gives rise to cycloidal skyrmionic configurations called Néel skyrmion. During the submission of our article, Chen *et al*[18] reported a r.t. skyrmion in a Fe|Ni system with a magnetic exchange layer which provides an effective magnetic field to stabilize the skyrmion phase, and Jiang *et al*[19] observed micron-size skyrmions at room temperature in a system composed of a single nanometer thick CoFeB film in contact with Ta film, resulting presumably of a small DM interaction. Our main goal in this article is to extend the generation of interface-induced skyrmions from monolayer thick magnetic films[15-16] to multilayers by stacking layers of magnetic metals and nonmagnetic heavy metals (*e.g.* Pt and Ir) so as to induce a much larger DMI at all the magnetic interfaces (Fig. 1a) allowing the observation of small skyrmions at room temperature, as shown hereafter. The advantages of such innovative multilayered systems are twofold. Firstly we anticipate that thermal stability of skyrmions can be greatly improved, simply because of the increase of their magnetic volume, as it turns out for our samples that the same magnetic texture extends vertically throughout the multilayer (exactly as for coupled domain walls in similar types of multilayers[20]). Secondly, the choice of two different heavy metals *A* and *B* sandwiching each magnetic layer (FM,

Fig. 1a) can lead to additive interfacial chiral interactions increasing the effective DMI of the magnetic layer if the two heavy materials induce interfacial chiral interactions of opposite symmetries and parallel **D** (see Fig. 1a).

The first important conclusion drawn from the analysis of our observations is that the presence of circular-shape domains in the 30-90 nm size range cannot be accounted for by dipolar interaction[21-24]. As our micromagnetic simulations also show that these small circular-shape domains can only be explained by the presence of large DMI imposing a winding number equal to one, we identify them as skyrmions. We will also show that both the size of the circular-shape domains and its field dependence are consistent with simulations of skyrmions. We note that the size of the skyrmions in this article, between 30 and 90 nm depending on the field, corresponds to the size that can be imaged with our experimental technique, but indeed we predicted[25] recently that, with the same DMI, even smaller sizes can be obtained by tuning the other parameters such as perpendicular magnetic anisotropy or the thickness of the ferromagnetic layers. The stabilization in multilayers and at room temperature of individual nanoscale magnetic skyrmions induced by large chiral interactions represents the most important result of this work. Given the important features of magnetic skyrmions associated with their topological character (small size, easier current-induced propagation, smaller sensitivity to defects, and so on), this advance represents a definite breakthrough on the route to highly integrated skyrmion-based memory devices[3].

**Multilayers with additive chiral interaction at interfaces with heavy metal layers**

The prototype of the multilayered systems that we studied is presented in Fig. 1a. The samples grown by sputtering are stacks of ten repetitions of an Ir|Co|Pt trilayer, each trilayer being composed of a 0.6-nm-thick Co layer sandwiched between 1 nm of Ir and 1 nm of Pt: Pt10|Co0.6|Pt1|{Ir1|Co0.6|Pt1}$_{10}$|Pt3 (numbers are thickness in nm). The choice of the two heavy materials *i.e.* Pt and Ir, has been guided by recent experiments of asymmetric domain wall propagation[26-27] as well as recent *ab initio* predictions of opposite DMI for Co on Ir and Co on Pt[28], which corresponds to additive DMI at the two interfaces of the Co layers sandwiched between Ir and Pt. In addition to these Ir|Co|Pt asymmetric multilayers, we also prepared reference samples of Pt|Co|Pt with symmetric interfaces, a type of structure investigated in previous studies which have shown that a non-complete cancellation of the DMI at the Pt/Co and Co/Pt may lead to a small but non-negligible global DMI[27-28]. Details about the growth conditions and the characterization of their magnetic properties are presented in the Methods section.

We present hereafter micromagnetic simulations[30-31] for the simplest assumption in which the experimental magnetization *M* and magnetic anisotropy *K* are distributed uniformly in the ten 0.6 nm thick Co layers and equal to zero elsewhere. The simulations are performed for one of these Co layers with the experimental values of *M* and *K*. Moreover, the potential impact of the exchange stiffness A of the magnetic Co layer is investigated by performing a series of micromagnetic simulations with different amplitude of *A*. Dipolar interactions are always included. In a second series of simulations (see Supplementary Material), to take into account that a part of the experimental values of *M* and *K* comes from the proximity-induced magnetism in Pt or Ir, we evaluate the maximum possible changes of our results induced by such spreading of the magnetization by assuming a dilution of *M, K* and *A* inside the multilayer. This leads to a slightly different estimate of the interfacial DMI magnitude without affecting our main conclusion.

**Magnetization mapping in asymmetric multilayers: Bubbles or skyrmions?**

The mapping of the out-of-plane component of the magnetization in our Ir|Co|Pt multilayers and its evolution as a function of the external perpendicular field have been performed by scanning transmission x-ray microscopy (STXM) on samples grown on $Si_3N_4$ membranes and measuring the dichroic signal at Co $L_3$ edge. In Fig. 1b-e, we display experimental images obtained at different $H_\perp$ values. After saturation at large negative field and inversion of the field, we first observe a domain configuration (Fig. 1b at 8 mT) that combines some labyrinth-shaped domains with other domains having almost a circular shape. When the field is increased to $\mu_0 H_\perp = 38$ mT, the magnetic domains favored by the field extend (red domains in Fig. 1c). Before the complete saturation of the magnetization is reached (see Fig. 1e at $\mu_0 H_\perp = 83$ mT), Fig. 1d (at 68 mT) is typical of a field range up to about 80 mT in which isolated magnetic domains of approximate circular shape persist in an almost totally perpendicularly polarized sample. Since the STXM probes the total thickness, these images correspond to an average of the magnetic configuration throughout the ten magnetic layers. Because we observe mainly two opposite contrast amplitude (corresponding to $m_z = \pm 1$), we conclude that these magnetic configurations, worms or circular-shape domains from Fig. 1b to 1d, are throughout the magnetic thickness of the multilayer[20].

We now focus on the dimension of these circular-shape domains as well as on their increase when $H_\perp$ is decreased. We will show that both are consistent with a large DMI value and the corresponding skyrmion modeling. As shown in Fig. 1f-g, we can determine precisely the size evolution of selected isolated nanoscale domains at decreasing fields from the analysis of the STXM images (see Supplementary Information). Results for the Ir|Co|Pt multilayers are presented as squares in Fig. 2. We find that the diameter of the circular-shape domains goes from about 30 nm at $\mu_0 H_\perp = 73$ mT to 80 nm at $\mu_0 H_\perp = 12$ mT. At lower field (closer to zero), the magnetic contrast evolves towards worm-like domains from which a proper diameter can no longer be defined. We emphasize that the diameter of the circular-shape domains (around 80 nm) that we observe at very low field values remains extremely small compared to usual values (at least larger than 800 nm with our films parameters) observed in classical bubble systems[21-23] in which the driving mechanism for bubble stabilization is the dipolar interaction.

To get more insights about the actual interaction at play in our multilayered Ir|Co|Pt systems, we compare in Fig. 2 the experimental data for the field dependence of the size of the circular-shape domains with micromagnetic simulations of DMI-induced skyrmions. It is first important to emphasize that, in these simulations using as inputs the experimental magnetization and anisotropy constant, dipolar interactions can never stabilize a classical bubble below typically the micrometer range. A second important issue for the determination of the DMI magnitude is to evaluate the impact of the value of the exchange constant $A$ in Co. $A$ is indeed particularly difficult to estimate for ultrathin magnetic layers and has not been obtained experimentally here. However it has been observed recently in similar stacking of Co thin films with heavy metals[32] that $A$ is markedly influenced the surface/volume ratio and evolves from 15 pJ/m for multilayers with 7 nm-thick Co layers (typical value for bulk Co) down to about 7 pJ/m for 2 nm-thick Co layers. Consequently we decided to compare the experimental results with three series of simulations obtained with different exchange constants ranging from 5 pJ/m to 15 pJ/m. The simulated diameters *vs* $H_\perp$ curves corresponding to different values of DMI are compiled in Fig. 2. It is first important to point out that, with the DMI taken into account and whatever the value of A, bubble-like configurations relax for $|\mathbf{D}| \geq 1$ mJ/m$^2$ in a stable configuration of isolated skyrmion having a winding number $|W|$ equals to 1, where $W = \frac{1}{4\pi} \int \mathbf{s} \cdot (\partial_x \mathbf{s} \partial_y \mathbf{s}) \mathrm{d}x\mathrm{d}y$, $\mathbf{s}$ being the normalized local magnetization (see *e.g.*

[24]). With the smallest exchange constant $A = 5$ pJ/m, isolated skyrmions are not stable anymore in our range of magnetic fields for |**D**| values larger than 1.6 pJ/m (see Fig. 2a). Thus we consider that a reasonable value for the actual exchange constant in our asymmetric multilayers is close to 10 pJ/m. For this intermediate value (see Fig. 2b), the best agreement with the experimental size and its variation with field is found for $|\mathbf{D}|= 1.9 \pm 0.2$ mJ/m$^2$. This large DMI is comparable, for example, with that derived recently in Pt|Co|AlO$_x$ trilayers from asymmetric domain nucleation experiments[33], domain wall annihilation[34], non-reciprocal spin wave propagation[35] or scanning nanomicroscopy[36]. This DMI is in good agreement with recent theoretical predictions adding the DMI magnitudes predicted for Pt|Co (3 monolayers) and Ir|Co (3 monolayers)[28]. Actually, this range of DMI values is indeed close to be optimal for the stabilization of isolated skyrmions (and so of isolated magnetic bits) instead of skyrmion lattices[25,37]. Finally, our approach is along the line of the experimental characterization of skyrmions developed by Romming *et al*[38], in which the evolution of the size and shape of skyrmions as a function of the magnetic field (using the model developed in [39]) is proposed to compare experimental spin polarized STM images with theory.

In addition to measurements on asymmetric multilayers, we performed a similar analysis with STXM images obtained on symmetric Pt|Co|Pt systems (see Supplementary Information). The best fit for the size and field dependence of circular-shaped domain configurations is obtained with DMI that are small but non-negligible. Indeed, it is known that the different structures of the two interfaces in Pt|Co|Pt-like trilayers induces a non-zero DMI[27,29].

To confirm the large DM magnitude in Ir|Co|Pt multilayers, we developed a second approach to estimate it from the analysis of the STXM images at remanence. This method relies on the quantitative analysis of the mean width of the perpendicular magnetic domains. From the images obtained at zero field after demagnetization shown as inset in Fig. 3, we can measure the average domain width to be $106 \pm 20$ nm for the Ir|Co|Pt multilayers. The same analysis of similar images for the symmetric Pt|Co|Pt multilayers leads to a mean domain width of $303 \pm 30$ nm. In Fig. 3, these experimental values are compared with those obtained in a series of micromagnetic simulations as a function of the DMI, allowing the estimation of *D*. The material parameters used in these simulations are the same as those used in Fig. 2b. From this second approach we find that the DMI is about $1.6 \pm 0.2$ mJ/m$^2$ for the asymmetric Ir|Co|Pt multilayers or $0.2 \pm 0.2$ mJ/m$^2$ for the symmetric Pt|Co|Pt sample, thus close to the values derived from the field dependence of the skyrmion size. Once again we emphasize that the domain width found in Pt|Co|Ir, in the 100 nm range at zero field, cannot be expected with only dipolar interactions[21].

As mentioned above, we also evaluated how our estimation of the DMI is modified if we take into account the proximity-induced magnetization of Ir and (Supplementary Information). We find that, in this case, the DM magnitudes from the asymmetric multilayers is about $1.4 \pm 0.3$ mJ/m$^2$, the actual value being probably between these two estimations. In consequence, all these results converge to the existence of very large interfacial DMI, and room temperature skyrmions in technologically relevant systems of magnetic multilayers.

**Room temperature observation of isolated skyrmion in magnetic disks**

As described in the previous section, we find from two independent analyses of the magnetic configurations that very large DMI and skyrmions exist at room temperature in asymmetric {Ir|Co|Pt}$_{10}$ multilayered films. Here we demonstrate that isolated nanoscale skyrmions can be stabilized at r.t. in nanodisks and nanostrips patterned in our multilayers by electron beam lithography and ion beam etching. In Fig. 4a, we show the field dependence of the diameter of an approximately circular domain located close to the center of a 500 nm-diameter disk and this dependence is compared, as we did in the first section, to what is obtained in micromagnetic simulations. Even if the agreement is less good than in extended films shown in Fig. 1 and Fig. 2, notably in the low field region, again a major outcome of these simulations is that it is not possible to stabilize any bubble-like domain in sub-micrometer-sized disks without introducing large DM values of at least 1.5 mJ/m$^2$. As the winding number of the simulated circular domains is, after stabilization, always equal to one, we can conclude that the experimental images correspond to nanoscale skyrmions with a chirality fixed by the sign of *D*. In the STXM images of Fig. 4b on 300 nm disks and 200 nm-wide tracks, we show even smaller skyrmions that are stable down to very small field (~8 mT) with dimensions ranging from 90 nm close to zero field to 50 nm in applied field. The fact that the observed skyrmion diameter does not depend significantly on the disk diameter is expected from our previous numerical study[25] in which it depends rather on the effective ratio between DM and exchange interactions as long as the DM magnitude remains smaller than the threshold value corresponding to negative domain wall energy. Note that, during the submission of this work and also in multilayers, Woo et al[40] have reported the room temperature observation of multiple skyrmions of relatively large diameter (around 400 nm) in 2-µm-diameter disks patterned in Pt|Co|Ta multilayers.

**Conclusion**

Our r.t. observation of individual sub-100 nm skyrmions that are produced in magnetic multilayers by a strong and additive interfacial chiral interaction represents the main achievement of this work. Ten repetitions of the Pt|Co|Ir unit are enough to stabilize firmly the skyrmions against thermal fluctuations at room temperature. We have already shown previously[25] that the size of such interface-induced skyrmions (down to 30 nm in the present series) could be reduced even more by tuning the magnetic anisotropy and described how they could be nucleated by spin injection and displaced by the spin Hall effect of the Pt layers. We believe that this experimental breakthrough can be a robust basis of the development of skyrmion-based devices for memory and/or logic applications, as well as the starting point of further fundamental studies on the very rich physics of skyrmions.

## Figures

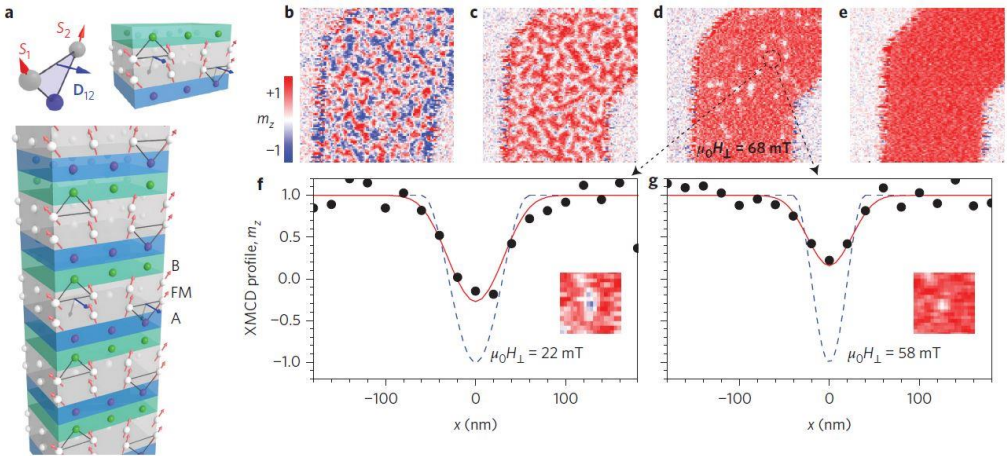

**Figure 1 | Interfacial DMI in asymmetric magnetic multilayers. a**, Top left: the DMI for two magnetic atoms (grey spheres) close to an atom with a large spin–orbit coupling (blue sphere) in the Fert–Levy picture14. Top right: zoom-in on a single trilayer composed of a magnetic layer (FM, grey) sandwiched between two different heavy metals A (blue) and B (green) that induce the same chirality (same orientation of **D**) when A is below and B above the magnetic layer. Bottom: a zoom-in on an asymmetric multilayer made of several repetitions of the trilayer. **b–e**, A 1.5 × 1.5 µm2 out-of-plane magnetization (mz) map obtained by STXM on a (Ir|Co|Pt)10 multilayer at r.t. for applied out-of-plane magnetic fields of 8 (**b**), 38 (**c**), 68 (**d**) and 83 (**e**) mT. **f**, Experimental X-ray magnetic circular dicroism (XMCD) signal through a magnetic circular domain (skyrmion) as observed at 22 mT (black dots). The blue dashed curve is the magnetization profile of an ideal 60-nm-diameter skyrmion and the red curve derives from the model described in the text. g, Same type of data at 58 mT and the corresponding simulation of a 40-nm-diameter skyrmion. The images and data of **f** and **g** result from the same skyrmion evolution in the field. The actual image size of the insets is 360 nm.

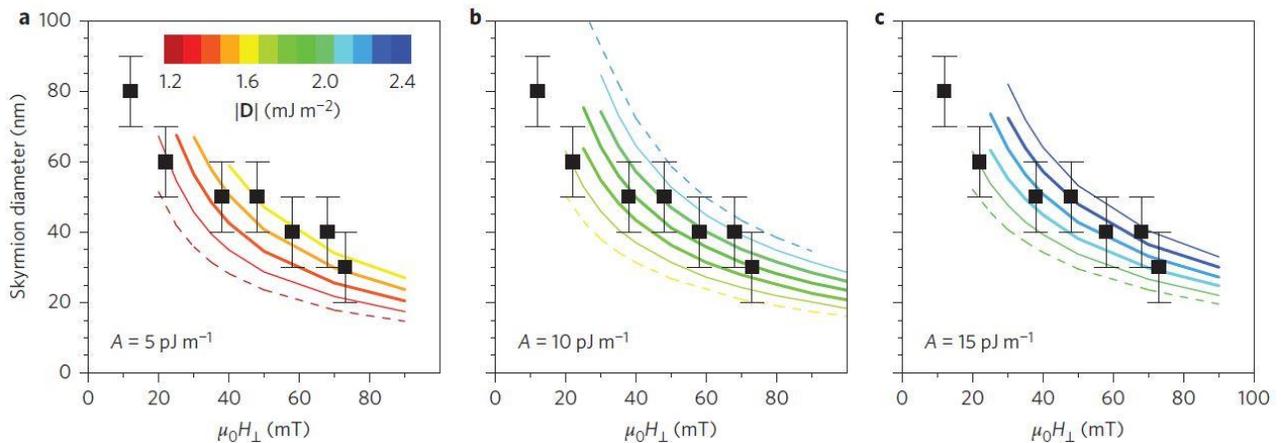

**Figure 2 | Skyrmion diameter as a function of the external out-of-plane magnetic field H⊥. a–c**, Experimental diameters (squares) can be compared with micromagnetic simulations for different values of |**D**| (lines) and different values of the exchange interaction A. Simulations were obtained in the case that gives a lower bound to the experimental |**D**| value (no exchange coupling between

magnetic layers). For A=5 pJ m−1, circular skyrmions are not stable for |**D**| ≥ 1.7 mJ m−2 and deform into labyrinthic worm-like structures. The deviation of the data compared with the simulations at a low field can be explained by the interaction with the other domains.

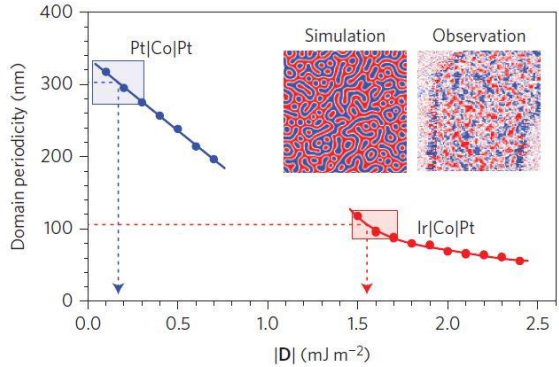 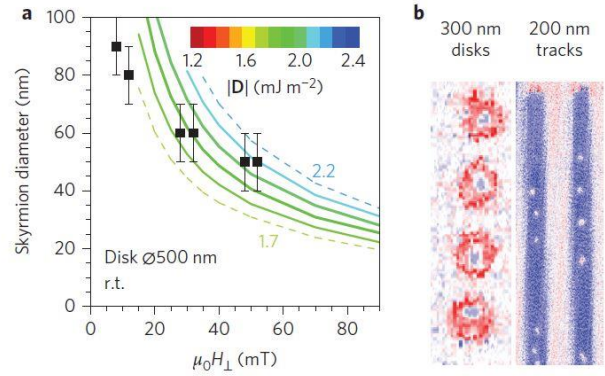

**Figure 3 | Micromagnetic simulations and experimental measurements of mean domain-width evolution with DMI after demagnetization.** Comparing the simulations with the experimental domain-width value (dotted horizontal line) allows us to estimate |**D**|(Ir|Co|Pt) = 1.8 ± 0.2 mJ m−2 and |**D**|(Pt|Co|Pt) = 0.2 ± 0.2 mJ m−2. The box height represents the error margins on the experimental domain-size evaluation; its width gives the resulting error on |**D**| for the used simulation parameters. The inset shows a simulated worm pattern for |**D**| = 1.6mJ m−2 in Ir|Co|Pt (1.5 × 1.5 μm2) and a corresponding experimental observation at the same scale (using the same colour code as in Fig. 1).

**Figure 4 | Evolution of the skyrmion size in patterned nanoscale disks and tracks. a**, Magnetic field evolution of the skyrmion size derived from micromagnetic simulations realized for A = 10 pJ m−1 (lines for different |**D**| values) and sizes of the observed skyrmions (squares) for 500-nm-diameter disks. **b**, R.t. out-of-plane magnetization map of an array of 300-nmdiameter disks with an out-of-plane external field of μ0H⊥ =8mT, and two 200-nm-wide tracks at μ0H⊥ ≈ 55 mT that display several isolated skyrmions.

**Acknowledgments:** The authors acknowledge technical support from Blagoj Sarafimov and Michael Bechtel for their technical support at the SLS and Bessy II beamlines. The STXM experiments were performed using the X07DA (PolLux) beamline at the Swiss Light Source, Paul Scherrer Institüt, Villigen, Switzerland and the Maxymus beamline BESSY II, Adlershof, Germany. The authors acknowledge financial support from the ANR agency, project ANR-14-CE26-0012 ULTRASKY and from the EU grant MAGicSky No. FET-Open-665095.

**Author contributions:** NR, CM, VC and AF conceived the project. CD and CML grew the films. CAFV, KG, NR patterned the samples. CM, CML, NR, JS, NVH, CAFV, KB, PW, PW, MW, JR and VC acquired the data at the synchrotron. CML and NR treated and analyzed the data with the help of CM, PW and VC. CML, JS and NR performed the micromagnetic simulations. CML, NR, VC and AF prepared the manuscript. All authors discussed and commented the manuscript.

**Additional information:** Supplementary information accompanies this paper at www.nature.com/naturenanotechnology. Reprints and permission information is available online at http://npg.nature.com/reprintsandpermissions/. Correspondence and requests for materials should be addressed to NR, CM or VC.

**Corresponding authors:** Nicolas Reyren (nicolas.reyren@thalesgroup.com); Christoforos Moutafis (christoforos.moutafis@manchester.ac.uk); Vincent Cros (vincent.cros@thalesgroup.com)

**Figure captions :**

**Figure 1 | Interfacial Dzyaloshinskii-Moriya interaction (DMI) in asymmetric magnetic multilayers. a**, The DMI for two magnetic atoms (gray spheres) close to an atom with large spin-orbit coupling (blue sphere) in the Fert-Levy picture[14]. Zoom on a single trilayer composed of a magnetic layer (FM, gray) sandwiched between two different heavy metals A (blue) and B (green) that induce the same chirality (same orientation of *D*) when A is below and B above the magnetic layer, and finally on an asymmetric multilayer made of several repetition of the trilayer. **b-e**, 1.5×1.5 µm$^2$ out-of-plane magnetization ($m_z$) map obtained by scanning transmission x-ray microscopy on a {Ir|Co|Pt}$_{10}$ multilayer at room temperature for applied out-of-plane magnetic fields of 8 (b), 38 (c), 68 (d) and 83 mT (e). **f**, Experimental dichroic signal through a magnetic circular domain (skyrmion) as observed at 22 mT (black dots). The blue dashed curve is the magnetization profile of an ideal 60 nm-diameter skyrmion and the red curve derives from the model described in the text. **g**, Same type of data at 58 mT and corresponding simulation of 40 nm-diameter skyrmion. Images and data of (f) and (g) panels are the result of the same skyrmion evolution in field. The actual image size of the insets is 360 nm.

**Figure 2 | Skyrmion diameter as a function of the external out-of-plane magnetic field $H_\perp$.** Experimental diameters (squares) can be compared to micromagnetic simulations for different values of *D* (lines) and different values of the exchange interaction *A* (panels). Simulations were obtained in the case which gives a lower bound to experimental *D* value (no exchange coupling between magnetic layers). Note that, for $A = 5$ pJ/m, circular skyrmions are not stable for $D \geq 1.7$ mJ/m$^2$ and deform into labyrinthic worm-like structures. The deviation of the data compared to the simulations at low field can be explained by the interaction with the other domains.

**Figure 3 | Micromagnetic simulations and experimental measurements of mean domain width evolution with DMI after demagnetization.** Comparing the simulations with the experimental domain width value (dotted horizontal line) allows to estimate: $D(Ir|Co|Pt) = 1.8 \pm 0.2$ mJ/m$^2$ and $D(Pt|Co|Pt) = 0.2 \pm 0.2$ mJ/m$^2$. The box height represents the error margins on the experimental domain size evaluation; its width gives the resulting error on *D* for the used simulation parameters. The inset shows a simulated worm pattern for $D = 1.6$ mJ/m$^2$ in Ir|Co|Pt (1.5×1.5 µm$^2$) and a corresponding experimental observation at the same scale (using the same color code than in Fig. 1).

**Figure 4 | Evolution of the skyrmion size in patterned nanoscale disks and tracks. a**, Magnetic field evolution of the skyrmion size derived from micromagnetic simulations realized for $A = 10$ pJ/m (lines for different *D* values) and sizes of the observed skyrmions (squares) for 500-nm-diameter disks. **b**, Room temperature out-of-plane magnetization map of an array of 300-nm-diameter disks with an out-of-plane external field $\mu_0 H_\perp = 8$ mT, and two 200-nm-wide tracks at $\mu_0 H_\perp \approx 55$ mT displaying several isolated skyrmions.

**Methods**

The studied samples are Co-based multilayers, which were grown by dc sputtering at room temperature under an Ar pressure of about 0.25 Pa on two different substrates: thermally oxidized Si wafer or 200-nm-thick $Si_3N_4$ membranes. The samples grown on $SiO_2$ were used to determine the macroscopic magnetic properties of the samples, while the ones on membranes were used for scanning transmission x-ray microscopy. A 10-nm-thick buffer made of textured Pt ((1 1 1)-oriented fcc) is first deposited, immediately followed by the growth of a Co(0.6 nm)|Pt(1 nm) bilayer. This first bilayer is used to induce a strong enough perpendicular magnetization anisotropy (PMA). We subsequently grew asymmetric structures {Ir(1 nm)|Co(0.6 nm)|Pt(1 nm)}$_{10}$ or symmetric ones {Co(0.6 nm)|Pt(1 nm)}$_{10}$. Both types of magnetic multilayers are finally capped by an extra 3-nm-thick Pt layer to prevent oxidation. The magnetic properties of the multilayers grown on $SiO_2$ were measured by SQUID or alternating gradient field magnetometry. The saturation magnetization and the effective anisotropy are determined from the magnetization loops with out-of-plane and in-plane magnetic fields. From these measurements, we deduce a saturation magnetization of $0.96 \pm 0.10$ MA/m ($1.6 \pm 0.2$ MA/m) and an effective anisotropy of $0.17 \pm 0.04$ MJ/m$^3$ ($0.25 \pm 0.07$ MJ/m$^3$) for the Ir|Co|Pt (Pt|Co|Pt) system. The configuration of their vertical magnetic domains has been also investigated by magnetic force microscopy (MFM) on the very same multilayers grown by sputtering deposition on $SiO_2$/Si substrate rather than on $Si_3N_4$ membranes. Importantly, we find similar values of the mean domain width for images recorded at zero field after a demagnetization process, meaning that the estimated DM magnitudes are equal to those determined by STXM.

The magnetic imaging presented here has been realized by scanning transmission x-ray experiments that have been performed on two different beamlines: X07DA (PolLux) beamline at the Swiss Light Source, Paul Scherrer Institute, Villigen, Switzerland, and the Maxymus beamline BESSY II, Adlershof, Germany. The images were recorded by scanning the multilayers grown on $Si_3N_4$ membranes with an x-ray beam focused by a Fresnel zone plate, providing a resolution down to 30 nm. Circular polarized light with normal incidence is used to map the out-of-plane magnetization at the nano-scale based on the x-ray magnetic circular dichroism (XMCD) effect (see Fig. S1-S3 in Supplementary Information). For imaging, we scan the samples at the Co $L_3$-edge at 779.95 eV. In order to get quantitative information about the DMI, we analyze the evolution of skyrmion dimensions with the out-of-plane field. The actual diameter of the circular-shaped domains is estimated by a fit of the experimental STXM data with the convolution of a modeled magnetization profile with the idealized Gaussian profile of the x-ray beam (as shown in Fig. 1e and 1f, and Fig. S2).

The quantitative estimation of the DMI has been obtained by comparison of the experimental results with micromagnetic simulations that have been performed using two different solvers, either the Object-Oriented MicroMagnetic Framework (OOMMF)[30] or MuMax3[31], both of which take into account the interfacial DM energy. We have checked that both solvers lead to strictly equivalent results. In order to perform these simulations, we have to introduce input material parameters that are the saturation magnetization $M_s = 956$ kA/m and the uniaxial out-of-plane magnetic anisotropy $K = 717$ kJ/m$^3$. The other parameter is the exchange constant $A$ that we take to be equal to $A = 10$ pJ/m if not stated otherwise. In the micromagnetic simulation of a skyrmion in perpendicular magnetic field, we relax the state at each magnetic field step starting from an initial state which is already a skyrmion. However, we check that, with our material properties, trivial bubbles (winding number equals zero) turn into skyrmions or vanish under high magnetic field ($\approx 50$ mT).